# CBM-Of-TRaCE: An Ontology-Driven Framework for the Improvement of Business Service Traceability, Consistency Management and Reusability

[1] Aida Erfanian, [2] Nima Karimpour Darav

*Islamic Azad University, Lahijan Branch, Iran*
[*1] erfanian@liau.ac.ir, [2] karimpour@liau.ac.ir

***Abstract.*** **In this paper, we represent a CBM-Of-TRaCE which is an ontological framework that integrates two aspects of business components: conceptual and methodology. In the development of our framework we have taken IBM's Actionable Business Approach (ABA) in to consideration. We evaluate our framework through some aspects such as support and facilitation for a business from five different aspects: service-orientation, business process, management integration, reusability improvement, consistency rules, and traceability. As well, we demonstrate the compatibility of our CBM-Of-TRaCE with ABA's four phases.**

*\*Corresponding address:*

Aida Erfanian,
*Islamic Azad University, Lahijan Branch, Iran*
 Email: *erfanian@liau.ac.ir*

## 1. Introduction

   Information technology offers a suitable infrastructure to manage the knowledge that flows through the business components. On the other hand, Component Business Model for the Business Of IT (CBMBoIT) has brought basic building blocks of an IT-based business within a two dimensional framework [7] which is namely coupled by Service-Oriented Architecture (SOA). One of the purposes of cooperation of this model with SOA is to guarantee the quality of business service [4]. However CBMBoIT clearly suffers from the lack of the explicit semantic relationships among its various components.

   Additionally, where and how can we express the track of SOA with the different components? on the other hand, IBM's Actionable Business Architecture (ABA) made the coordination between SOA and CBM more applicable [9]. ABA is a methodology, like any other methodology it could be more effective if it is supported by an appropriate infrastructure which leverages its benefits.

   Furthermore, as it is emphasized in [19], the documentation of an SOA plays an important role in a software-intensive system which doing it effectively requires years of experience from the past. Therefore, it implies that appropriate knowledge management is the key to fulfill this need. Knowledge management improves a business and the quality of its provided services from different aspects [18]. Through the past years ontologies have proved their strong role in domains like enterprise knowledge management [13]. Also ontologies facilitate storage and organization of relationships among business processes [5].

That is the reason why we propose an ontological framework based on CBM, BPM, and SOA suitable for ABA with some advantages, which can accumulate the knowledge in an effective way. In other words, handling the variety of dependencies such as consistency management among different building blocks of a normal CBM framework require a rough checking process which probably takes lots of time and effort. ABA, as a coordinator between SOA and CBM, solves this problem to some extents, but it does not provide a clear systematic infrastructure to gather the variety of knowledge gained via its different stages. Hence, based on ABA, we proposed our framework as an effective infrastructure which addresses this issue as well as other advantages it brings such as the traceability, reusability, and consistency evaluation of the organized knowledge of the business.

Vice versa, via a semantic checker or a tool which can keep the track of the components models, functions and value chain, imposing controllable explicit semantics and their rules makes the evaluation process of a business easier. In addition, business knowledge which is maintained and well-organized through an ontology is more reusable and traceable with the potential of consistency management by well-defined semantic rules. The main aim of this paper is to resolve the mentioned issues above and represent a solution by an ontological framework which is called CBM Ontology for the improvement of Traceability, Reusability and Consistency Evaluation (CBM-Of-TRaCE). As it is going to be shown later in this paper, by the means of CBM-Of-TRaCE, a better understanding of processes and their deficiencies is provided for a business in which reusability, traceability and consistency-checking among the components are the major staple necessities.

Various approaches for the documentation and organization of SOA and Business Models have been proposed by different researchers through the recent years. A.Osterwalder and Y.Pigneur in"An e-Business Model Ontology for Modeling e-Business" [15], M.Korotkiy and J.L.Top in"Onto-SOA: From Ontology-enabled SOA to Service-enabled Ontologies" [12] and B.Andersson et al. in "Towards a Reference Ontology for Business Models" [2] represent ontologies for SOA and Business. But, none of them combine SOA and Business ideas in an ontology based on CBM which makes our work different from theirs. Although B.M.Bidgoli and L.Rafati in"Business Service Modeling in Service-Oriented Enterprises (SOBM)" [14] made such an effort but their focus is only on the binding of CBM to SOA. Nonetheless, we extended and optimized their work to coordinate it with the IBM's Actionable Business approach [9] which gathers the best practices like CBM map, SOMA method, and BPM together to make the most of them. In addition to more conformance with ABA we focused on our framework to provide a foundation for more reusability and rules for the evaluation for consistency of a service-oriented business.

In this paper, we represent our framework which we named CBM-Of-TRaCE (which stands for CBM Ontology for the improvement of Traceability, Reusability and Consistency Evaluation) in section 3. Later, in section 3 we evaluate our framework as well. This evaluation is made by five criteria through which we try to show that, comparing to other business models such as CBM and SOBM, our framework is beneficial by considering different aspects. These criteria include Support for SOA, BPM Integration, Reusability Improvement, Consistency-Rules, and Traceability. Actionable Business Architecture (ABA) is a proven method that plays an important role towards coordination of CBM and SOA, so we did our best to make our framework compatible with it. Therefore we present the compatibility of our framework with ABA in section 4. Afterwards, we sum up our paper and discuss future works, open issues and challenges as a conclusion represented in section 5.

## 2. Literature Review

To clarify the idea behind our proposed framework we need to explain some literature that we used as a basis.

### 2.1. Component Business Model

In this section we represent Component Business Model (CBM) based on [7] and [16]. Organizations can use the concept of CBM in order to realize their proficiency. CBM enables organizations to evaluate their goals and strategies as well as reducing the risks without any complexity.

#### 2.1.1. *Business Components*

The business components are the modular building blocks of an organization. As it is demonstrated in Figure 1, each component has five dimensions: business purpose, activities, resources, governance, and business services. Business purpose is a rationale which describes the existence of a business component. Each component does a set of activities to reach its purpose. Components need resources which includes people, knowledge, and assets. On the other hand, activities are supported by resources. Based on the governance model, each component is managed by an independent entity. Moreover, like each business, each component provides a number of business services.

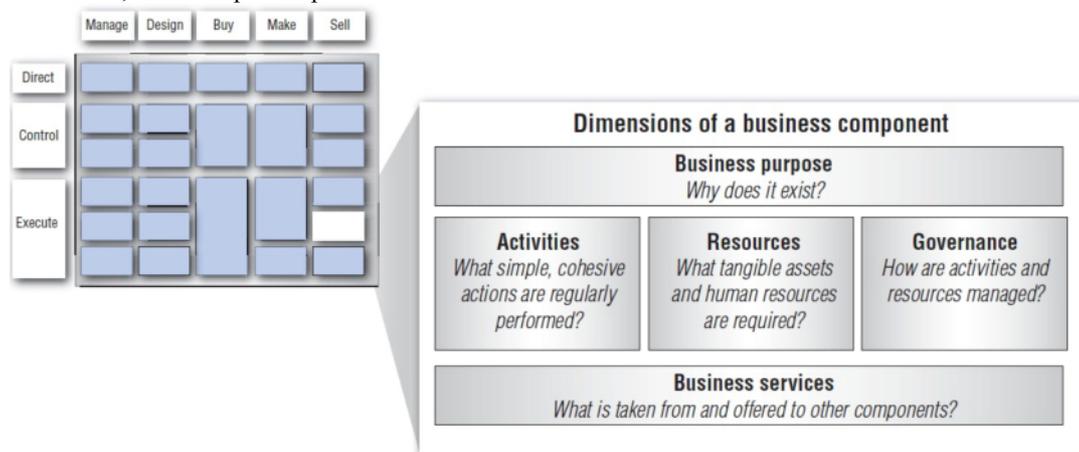

**Figure 1**: Dimensions of a business component within CBM [16].

### 2.2. Actionable Business Architecture

In this section we represent Actionable Business Architecture (ABA) based on [9]. Today, executive managers keep thinking about innovation in their strategies in order to sustain the development of their business. Realization of business strategies and the incremental growth of business values require effective and mutual relationships among strategy, operation and Information Technology (IT). ABA defines the relationships and collaborations among different domains such as strategy, operation, and IT models and enables the removal of the existing gap between initial goals and results, as well as prevention of the loss of opportunities and acceleration in time-to-value.

ABA is consisted of models, methods, metrics, and tools. Models are reusable assets, standards, and directives of the knowledge of the business. Methods are the techniques that are applied in different phases. These phases are called business insight, operation design, service identification, and business investment. Metrics are the keys of performance, agility, and risk markers which are mainly used to measure the goals. Tools facilitate the application of models, methods, and metrics.

### 3. Our Proposed Framework

CBM Ontology for the improvement of Traceability, Reusability and Consistency Evaluation (CBM-Of- TRaCE) is suggested to address the deficiencies of Component business Model (CBM) which is tried to be alleviated by a method proposed by IBM called Actionable Business Architecture (ABA). ABA is a successful method and practically proved, but still suffers from the lack of the organization of knowledge which could be collected and managed more efficiently if it was supported by an effective infrastructure such as CBM-Of-TRaCE.

Therefore, our framework is able to solve this problem by settling knowledge and maintain meaningful relationships among different components of a business. Knowledge management in a business which uses CBM as a model can be done more effectively if our CBM-Of-TRaCE is used within since it improves staple necessities of knowledge management such as reusability, traceability, consistency evaluation and SOA and BPM as well. CBM-Of- TRaCE is a very high-level ontology for the CBM [16] which can then be synthesized in to lower levels. Therefore, we can consider it as the first-level ontology. Each of the components of the first level can be decomposed in to a lower-level which is comprised of its sub-components and their semantically expressed relations.

We should also take the maintenance of the balance and consistency among the different levels in to consideration. This issue can be resolved by defining a meaningful relationship between the adjacent levels either. This issue needs more effort so it is considered for future work. CBM-Of-TRaCE is demonstrated in Figure 2.

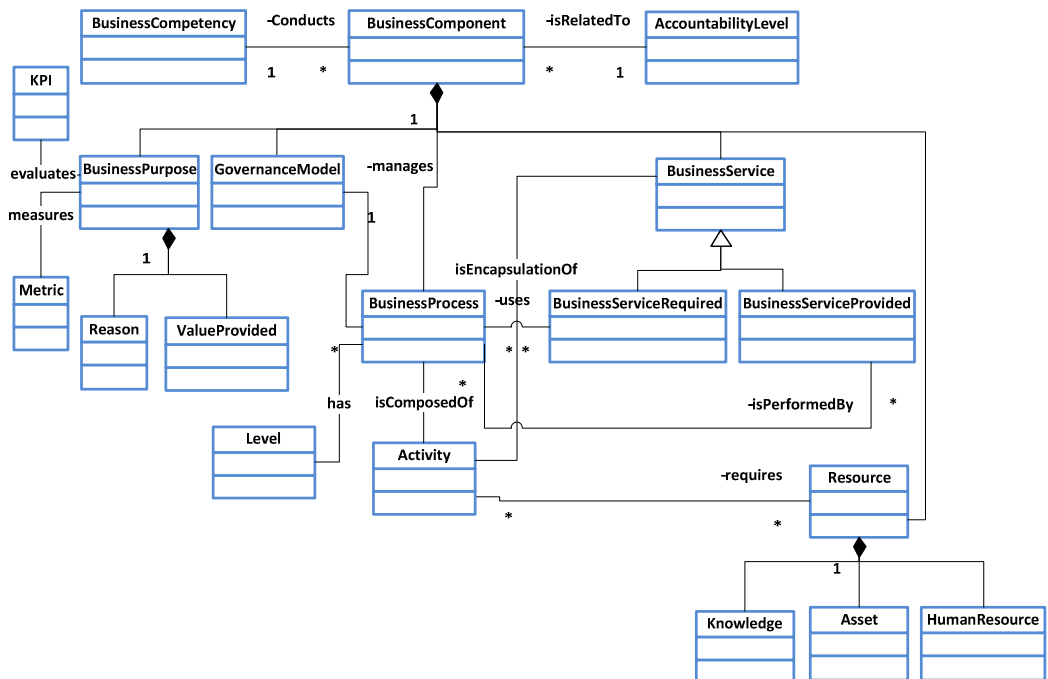

**Figure 2**: The CBM-Of-TRaCE schema which is consisted of a business component and its relationship with CBM-framework dimensions with respect to IBM's SOMA and ABA methods

As Figure 2 depicts, each business component is related to an accountability level and is conducted by a business competency. Besides, each business component is composed of a business purpose, governance model, some business services, and resources. A governance model also manages some activities. A business purpose itself, is composed of a reason and it provides value to the business. Additionally, each business service can be classified as a business service that is required and a business service that is provided to other business components. Furthermore it is performed by some activities. An activity requires number of resources which includes knowledge, asset, and human resource.

A business process, which resides in a level, is composed of activities that are encapsulation of business services. Moreover, a business purpose can be evaluated by the means of key performance metrics (KPIs) and measured by other metrics.

### 3.1. Evaluation of Our Framework

In this section we are going to compare and evaluate our proposed framework with the SOBM and CBM via five criteria: Support for SOA, BPM Integration, Reusability Improvement, Consistency-Rules, and Traceability.

#### 3.1.1. *Support for SOA*

In CBM [7] it is stated that the business components provide business services to each other which one can obviously conclude that CBM supports Service Orientation. As it is insisted and shown in SOBM [14], SOBM is clearly designed for Service-Orientation. From this aspect our proposed framework supports service-orientation, too. However, we explicitly expressed and revealed the implicit interaction among business components via our proposed ontology. That is why one can find the business service class below the business component one in a composition relationship in Figure 2. Although CBM, SOBM, and CBM-Of-TRaCE support service orientation, CBM-Of-TRaCE simplified and provided a way to model the business services via the defined ontology which can be counted as one of the trues virtues of our technique.

#### 3.1.2. *BPM Integration*

Business Processes and their analysis can done more efficient if they are expressed by models, especially when they are formal [20]. Our CBM-Of-TRaCE model is a place to define the processes since we depicted the relationship among them. Moreover, because of its ontological nature and the ability to define constraints such as consistency rules it brings the required formality. Hence, we can conclude that our framework integrates the BPM. But, because of no business process existence in SOBM, there is not any clear consideration of the association between components, services, and processes in SOBM. Nevertheless, as demonstrated in SOMA [3] through the images, business processes lay behind the idea of CBM in each component without a clear explicit presentation in the CBM framework.

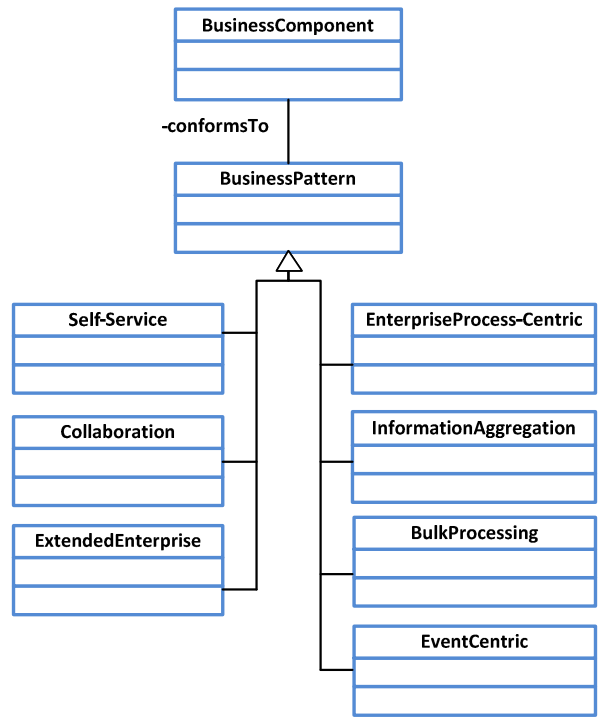

**Figure 3:** Integrating patterns into our proposed framework

### 3.1.3. *Reusability Improvement*

To simplify and provide more reusability to our framework we decided to extend CBM-Of-TRaCE to make the use of business patterns. As Adams et al. in [1] discussed, their proposed business patterns provide reusability. Therefore, we found our framework a perfect place to consider their patterns: Self-Service, Collaboration, Information Aggregation, Event-Centric, Enterprise Process-Centric, Extended Enterprise, and Bulk Processing. CBM-Of-TRaCE already couples with CBM and the idea of the separation of concerns [8] because we built it upon these concepts and with keeping an eye on IBM's patterns for e-business expressed in [11]. Our proposed ontology provides a way to define reuse-oriented associations between the best-experiences and the current business components.

In order to integrate these patterns into our framework and to sustain reusability we can insert a Business Pattern class in our ontology with a "conforms to" relationship that associates it with the Business Component class. In other words, every Business Component conforms to a Business Pattern. This is presented in Figure 3. Besides, neither of CBM or SOBM sustains a place for the application of patterns in order of more reusability and routing of each component to its root, the pattern.

### 3.1.4. *Consistency-Rules*

One of the advantages of the definition of an ontology for a framework is the formality which it brings to some extents [6]. Hence, formality provides us with automatic consistency checking by the means of the definition of meta-rules. Meta-rules are static and can be described as the set of rules which are defined as the essential part of the framework. Hence, no matter what is a model, whenever a model is constructed based on the ontological framework, it is automatically checked by a rule engine for eligibility. Furthermore, dynamic environmental rules are definable based on the directives and constraints of a business. Basically, CBM is not an ontology so there is no place for automation and it requires a proper tool in support. Likewise, the authors of SOBM did not considered any consistency rules conformance for their framework.

### 3.1.5. *Traceability*

Our proposed framework not only provides traceability among different business components and their sub-components but it provides the traceability between the business and corresponding patterns. Additionally, One can keep the track of the current business with the previous versions which is very beneficial to the cause and effect analysis of a business because all of the events and modifications on the road to an optimum solution can be basically maintained and managed in a knowledge repository which is built upon the CBM-Of-TRaCE.

### 3.1.6. *Evaluation Results*

As it is demonstrated above we can sum up our evaluation results from the discussed aspects mentioned in this section in to Table 1.

Table 1: Comparison of CBM-Of-TRaCE with SOBM and CBM frameworks

| *Business Model* | *Support for SOA* | *BPM Integration* | *Reusability Improvement* | *Consistency-Rules* | *Traceability* |
|---|---|---|---|---|---|
| CBM | Yes | Not clearly | No | No | No |
| SOBM | Yes | Not clearly | No | No | Limited |
| CBM-Of-TRaCE | Yes | Yes | Yes | Yes | Yes |

## 4. Compatibility with Actionable Business Architecture

Actionable Business Architecture (ABA) method [9] is an IBM's approach to unify and integrate separated set of methods such as CBM Method, Business Process Management (BPM Method) [10], and Service-Oriented Modeling and Architecture (SOMA Method) [3]. ABA is a proven method that leverages the need for reusability, traceability and efficiency with in a business [9] CBM-Of-TRaCE is highly compatible with ABA methodology since it covers all of the aspects of IBM's insight into its different phases: Business Insight, Service Identification, Operations Design, and Business Investment where each of these phases produces reusable deliverables [9].

However, what is absolutely necessary is an appropriate infrastructure to keep track of the transformations between phases, the processes within them and also the artifacts together. Nonetheless, the consistency between the products of each phase is the bottom line. That is why we constructed the CBM-Of-TRaCE through which all of the requisites above are fulfilled based on the facts we discussed in Section 2.1.

We claim that our framework is compatible with the ABA phases and facilitates them based on the following justification for each of the phases mentioned in [9]:

• Support for Business Insight:
  o By the means of the Business Purpose class with in our proposed ontology we can keep track of business goals with corresponding Business Components.
  o Once the Key Performance Metrics (KPIs), and other Metrics of the Business purpose are identified they can be resided in the ontology as KPI and Metric individuals or instances. Moreover, they can be used further in the analysis phase where the performance pain points have to be identified. This is invaluable as the

     knowledge which is gathered for future can be reused for optimization steps or cause and effect analysis.
- Support for Service Identification:
  - Because of the existence of the classes such as Business Component, Business Service, and Business Purpose and their associations in our proposed ontology, selection and insertion of information about the new and current business capabilities are simplified.
  - We have considered the interaction among business components via business services within our ontology.
- Support for Operations Design:
  - Having a glance over Figure 2, it is crystal clear that the mapping from the business processes to business components are indicated explicitly.
  - As long as our ontology is a formal model we can express and legislate any ground rules that are counted as meta-rules in addition to definition of rules that are based on business directives and constraints. This empowers the automation and consistency checking.
- Support for Business Investment:
  - We have considered "Asset class" as a type of Resource. This asset can be a software asset that a business component is realized by. Hence, Because of its conformance with ABA, the results of the evaluation criteria mentioned in Section3.1 are naturally boosted up. On the other hand, as long as it conforms well to ABA it can be adopted by a business which selected ABA as the methodology.

## 5. Conclusions

In addition to the support, integrity, and semantics the CBM-Of-TRaCE provides for the aspects discussed above (SOA, BPM, reusability improvement within a business, consistency rules definition ability, and traceability) it also can facilitate the application of CBM (Component Business Model) within an IT-based business from the following aspects:

- Making meaningful changes because, it is an ontology and it is a well-formed infrastructure with semantics by nature.
- Distinguishing between the components and values they bring, since the business components, their goals, and KPIs, and other metrics and their relation with processes and resources are well-defined within CBM-Of-TRaCE.
- Identification of opportunities in order to enhance the efficiency. Because it conforms to ABA and its phases, all of the evaluation and analysis results can be managed semantically and will be available with associations for further reuse such as the identification of opportunities.

  CBM-Of-TRaCE has the potential for tool support but in our current step of our work we concentrated only on the conceptual framework and the support for methodology. We chose ABA methodology over other methodologies because based on IBM reports it is well-proven and used in numerous successful projects. Therefore we focused our endeavor to construct an appropriate framework to provide an infrastructure with the potential of automation in some extents for businesses that have chosen ABA over other methodologies.
  In this paper we limited our concentration on the model and framework itself so we did not suggest any consistency rules. Our future work will be the development of a tool as well as some more ground integrity rules definition. Although, we demonstrated that our framework is applicable and effective from different aspects there all still open issues and challenges that need to be solved to make it more practical. These issues include the following items:

- Extending our model in to lower levels. For example, we can introduce a finer level for business process based on an appropriate meta-model presented by M.L.Rosa, et al. in [17].
- When refining the models into lower levels, the "balance" between adjacent levels should be defined

## References

[1] J. Adams, S. Koushik, G. Galambos, and G. Vasudeva. Patterns for e-business: A Strategy for Reuse. IBM Press, 2001.
[2] B. Andersson, M. Bergholtz, A. Edirisuriya, T. Ilayperuma, P. Johannesson, J. Gordijn, B. Gr´egoire, M. Schmitt, E. Dubois, S. Abels, A. Hahn, B. Wangler, and H. Weigand. Towards a reference ontology for business models. Conceptual Modeling - ER 2006, pages 482–496, 2006.
[3] Arsanjani and A. Allam. Service-oriented modeling and architecture for realization of an soa. In Proceedings of the IEEE International Conference on Services Computing, SCC '06, pages 5-21, Washington, DC, USA, 2006. IEEE Computer Society.
[4] F. Arslan. "Service oriented paradigm for massive multiplayer online games". International Journal of Soft Computing And Software Engineering, vol.2, No.5, pages 37–47, May 2012.
[5] R. M. Dijkman, M. L. Rosa, and H. A. Reijers. "Managing large collections of business process models-current techniques and challenges". Computers in Industry, 63(2):91–97, 2012.
[6] L. Elst and A. Abecker. Ontologies for information management: Balancing formality, stability, and sharing scope. Expert Systems with Applications, 23(4):357–366, 2002.
[7] M. Ernest and J. M. Nisavic. Adding value to the it organization with the component business model. IBM Syst. J., 46:387–403, March 2007.
[8] E. Ernst. Separation of concerns. In Proceedings of Software Engineering Properties of Languages for Aspect Technologies, SPLAT 2003, in assoc. with AOSD 2003, page 6 pages, 2003.
[9] R. Harishankar and S. K. Daley. Actionable business architecture. In Proceedings of the 2011 IEEE 13th Conference on Commerce and Enterprise Computing, CEC '11, pages 318–324, Washington, DC, USA, 2011.IEEE Computer Society.
[10] IBM-Corporation. Business process management. http://www-935.ibm.com/services/us/en/businessservices/bpm-enabled-by-soa-services.html.
[11] IBM-Corporation. Patterns for e-business for new and enhanced it solutions.
[12] M. Korotkiy and J. L. Top. Onto-soa: From ontology-enabled soa to service-enabled ontologies. In AICT/ICIW, page 124, 2006.
[13] A. Maedche, B. Motik, L. Stojanovic, R. Studer, and R. Volz. Ontologies for enterprise knowledge management. Intelligent Systems, IEEE [see also IEEE Intelligent Systems and Their Applications], 18(2):26–33, 2003.
[14] Minaei-Bidgoli and L. Rafati. Business service modeling in service-oriented enterprises. In Proceedings of the 2008 Fourth International Conference on Networked Computing and Advanced Information Management - Volume 02, NCM '08, pages 296–301, Washington, DC, USA, 2008. IEEE Computer Society.
[15] A. Osterwalder and Y. Pigneur. An e-business model ontology for modeling e-business. Industrial Organization 0202004, EconWPA, Feb. 2002.
[16] P. Pohle, G. Korsten and S. Ramamurthy. Component Business Models: Making Specialization Real. IBM Institute for Business Value, 2005.
[17] S. Lusa, D. I Sensuse. "Study of Socio-Technical For Implemetation of Knowledge Management System", International Journal of Soft Computing and Software Engineering [JSCSE], Vol. 2, No. 1, pp. 1-10, 2012.
[18] M. L. Rosa, H. A. Reijers, W. M. P. van der Aalst, R. M. Dijkman, J. Mendling, M. Dumas, and L. Garc´ıa-Ba˜nuelos. Apromore: An advanced process model repository. Expert Syst. Appl., 38(6):7029–7040, 2011.

[19] S. Tilley and S. Bellomo. 7th international workshop on graphical documentation: documenting soa-based systems. In Proceedings of the 27th ACM international conference on Design of communication, SIGDOC '09, pages 309–310, New York, NY, USA, 2009. ACM.

[20] M. P. van der Aalst, A. H. M. ter Hofstede, and M. Weske. Business process management: A survey. In Business Process Management, pages 1–12, 2003.

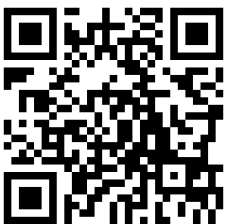

Free download and more information for this paper